# A magnetically collimated jet from an evolved star

Wouter H.T. Vlemmings*, Philip J. Diamond* & Hiroshi Imai**

*Jodrell Bank Observatory, University of Manchester, Macclesfield, Cheshire SK11 9DL, UK

**Department of Physics, Faculty of Science, Kagoshima University, Kagoshima 890-0065, Japan

**Planetary nebulae often have asymmetric shapes, which could arise due to collimated jets from evolved stars before evolution to the planetary nebula phase[1-3]. The source of jet collimation in these stars is unknown. Magnetic fields are thought to collimate outflows that are observed in many other astrophysical sources, such as active galactic nuclei[4-7] and proto-stars[8,9], although hitherto there are no direct observations of both the magnetic field direction and strength in any collimated jet. Theoretical models have shown that magnetic fields could also be the dominant source of collimation of jet in evolved stars[10,11]. Here we report measurements of the polarization of water vapour masers that trace the precessing jet emanating from the asymptotic giant branch star W43A at 2.6 kpc from the Sun, which is undergoing rapid evolution into a planetary nebula[2,12]. The masers occur in two clusters at opposing tips of the jets, ~1,000 AU from the star. We find direct evidence that the magnetic field is collimating the jet.**

The 22 GHz $H_2O$ masers can be observed at high angular resolution and can be used to trace the magnetic field strength and structure at small scales[13-15]. Circular polarization is caused by the Zeeman effect, observations of which yield the magnetic field strength along the line of sight[13,14,16]. The linear polarization vectors describe the direction of the magnetic field.

We have used the Very Long Baseline Array (VLBA) of the National Radio Astronomy Observatory to determine both the linear and circular polarization of the 22 GHz $H_2O$ masers in the jet of W43A (Figs 1 and 2). In our linear polarization spectra we achieve a r.m.s. noise of approximately 10 mJy beam$^{-1}$, and we detect linear polarization in the 6 brightest maser features (peak flux > 6 Jy beam$^{-1}$), with a weighted mean linear polarization fraction of 0.66 ± 0.07%. The 3σ r.m.s. upper limits of the linear polarization fraction on the other 14 detected but weaker features range upwards from 0.68%. The linear polarization vectors are predominantly aligned with the jet. According to maser theory, the linear polarization vectors are either parallel or perpendicular to the magnetic field direction[17]. The strongest linear polarization is typically found when the polarization vector is perpendicular to the magnetic field direction on the sky[13, 16]. We thus conclude that most of the observed linear polarization vectors are perpendicular to the magnetic field direction as shown in panel **c** and **d** of Fig. 1. Across the brightest maser feature we observe a 90° flip of the polarization angle. Here, as predicted by the theory and as previously observed in SiO masers[18], the linear polarization vector changes from parallel to perpendicular with respect to the magnetic field across the maser. We find the median, error weighted, position angle of the magnetic field to be -27 ± 12º east of north. This error estimate includes the systematic error (8º) introduced after polarization calibration with respect to the calibrator J1743-0350. The magnetic field direction is thus almost perfectly perpendicular to the jet, which is consistent with a toroidal magnetic field configuration. We have also detected a fractional circular polarization $P_V = 0.33 ± 0.09\%$ in one of the maser features in the southern tip of the collimated jet (Fig. 2). Using a full 22 GHz $H_2O$ maser radiative transfer code, that does not assume thermal equilibrium, to fit to the circular polarization[13-16], we find the magnetic field strength along the line of sight in the maser $B_\parallel = B \cos \theta = 85 ± 33$ mG, where $\theta$ is the angle between the line of sight and the magnetic field direction. We estimate $\theta \approx 65º$ using the linear polarization fraction (0.64 ± 0.19 %) of the maser and our polarization model fit results. Thus, the magnetic field in the $H_2O$ masers at one position within the collimated jet is $B \approx 200 \pm 75$ mG. This implies that, depending on the hydrogen number density in the $H_2O$ maser regions of the jet, the magnetic pressure dominates the gas pressure by a factor of 2-200, as expected in the case of magnetic collimation[11].

$H_2O$ masers at 22 GHz are excited in gas with a hydrogen number density $n \approx 10^8$-$10^{10}$ cm$^{-3}$ [19]. The $H_2O$ masers in the collimated jet of W43A likely arise when the jet has swept up enough material previously expelled from the star so that conditions at the tip of the jet have become favourable for $H_2O$ masers to occur[2]. As the typical hydrogen number density at 1,000 AU from the star is $n \approx 10^5$ cm$^{-3}$, the swept up material has increased the density in the maser region by a factor between ~$10^3$-$10^5$. When the magnetic field is partially coupled to the gas, compression of the magnetic field lines increases the field strength in the maser region, with the relation between the magnetic field strength and density $B \propto n^k$. Theoretical and observational considerations indicate that the coefficient $k$ can range between 0.3 and 1. An empirical relation, which has $k = 0.47$, was determined from observations of magnetic fields in star-forming regions[20]. Using this relation, the magnetic field strength in the lower density region outside the jet is $B \approx 0.9$-2.6 mG. This is consistent with values measured in the OH maser region at similar distances from the star in other systems. Alternatively, the masers may occur in a shock, similar to the $H_2O$ masers found in star-forming regions[21]. Such shocks might exist between the collimated jet and dense material in the outer circumstellar envelope. For a shock model we calculate, using $H_2O$ shock excitation models[21], that the pre-shock hydrogen density is ~3 $10^6$ cm$^{-3}$ and the pre-shock magnetic field strength is $B \approx 0.07$ mG.

A rotating magnetic field in the outflow of a star can be described by two components: a toroidal component $B_\varphi$ ($\propto r^{-1}$) and a radial component $B_r$ ($\propto r^{-2}$). Due to the dependencies on distance ($r$) from the star the radial component can be effectively neglected at distances of 1,000 AU. This is consistent with the observed magnetic field directions, which are fully toroidal. Using the $r^{-1}$ dependence, we can estimate the magnetic field strength $B_{\varphi s}$ at the base of the collimated jet on the surface of the star ($R \approx 1$ AU) to be ~1.5 G if the jet $H_2O$ masers are excited in swept up material, or ~70 mG if the $H_2O$ masers occur in shocks. The stellar toroidal magnetic field strength has a latitude dependence of the form $B_{\varphi s} (\theta_s) = B_s \sin \theta_s$, where $B_s$ is the strength at the equator $\theta_s = \pi/2$. Assuming we measure the magnetic field at the edge of the collimated jet with an opening angle of ~5°, this implies $B_s \sim 35$ or 1.6 G. The higher value is in excellent agreement with measurements extrapolated from previous maser polarization observations[14, 15, 18, 22].

The magnetic field and jet characteristics of W43A support the recent theoretical models that use magnetically collimated jets from evolved stars to explain the shaping of asymmetric planetary nebulae[11, 23, 24]. These models also explain the shapes of a large number of other proto-planetary nebulae, such as He 3-401[25]. However, the origin of the stellar magnetic field is still a matter of debate. Magnetic dynamo models invoking the differential rotation between a rapidly rotating core and a more slowly rotating outer layer have been shown to be able to produce magnetic fields in evolved stars similar in strength to those found in our observations[10]. Other models stress the need for a binary companion or heavy planet for a sufficient magnetic field to be attained[26]. The presence of a heavy planet in orbit around W43A could also be the cause of the observed jet precession[2, 27].

Different models have described the collimation of jets by magnetic fields around AGN[5] and proto-stars[28]. These models have been shown to be able to produce many of the observed characteristics of sources, such as the collimated proto-star HH212[9] and a large sample of active galactic nuclei (AGN)[6]. Additionally, recent laboratory work has been able to produce jets collimated by toroidal magnetic fields[29]. For the jets produced by AGN there are many indications that magnetic collimation is occurring[6, 7, 30], however, no observations yet exist of the magnetic field strength. There is also no direct evidence of magnetic collimation of proto-stellar jets. The characteristics of the jet of W43A are similar to those found in star-forming environments[9], and as magnetic fields of similar strengths are found in those regions[13], magnetic collimation such as described here likely also occurs during star-formation.

Acknowledgements: NRAO is a facility of the National Science Foundation, operated under cooperative agreement by Associated Universities, Inc. W.V. was financially supported by a Marie-Curie Intra-European fellowship.

Competing interests statement: The authors declare that they have no competing financial interests.

Correspondence and requests for materials should be addressed to W.V. (e-mail: wouter@jb.man.ac.uk).

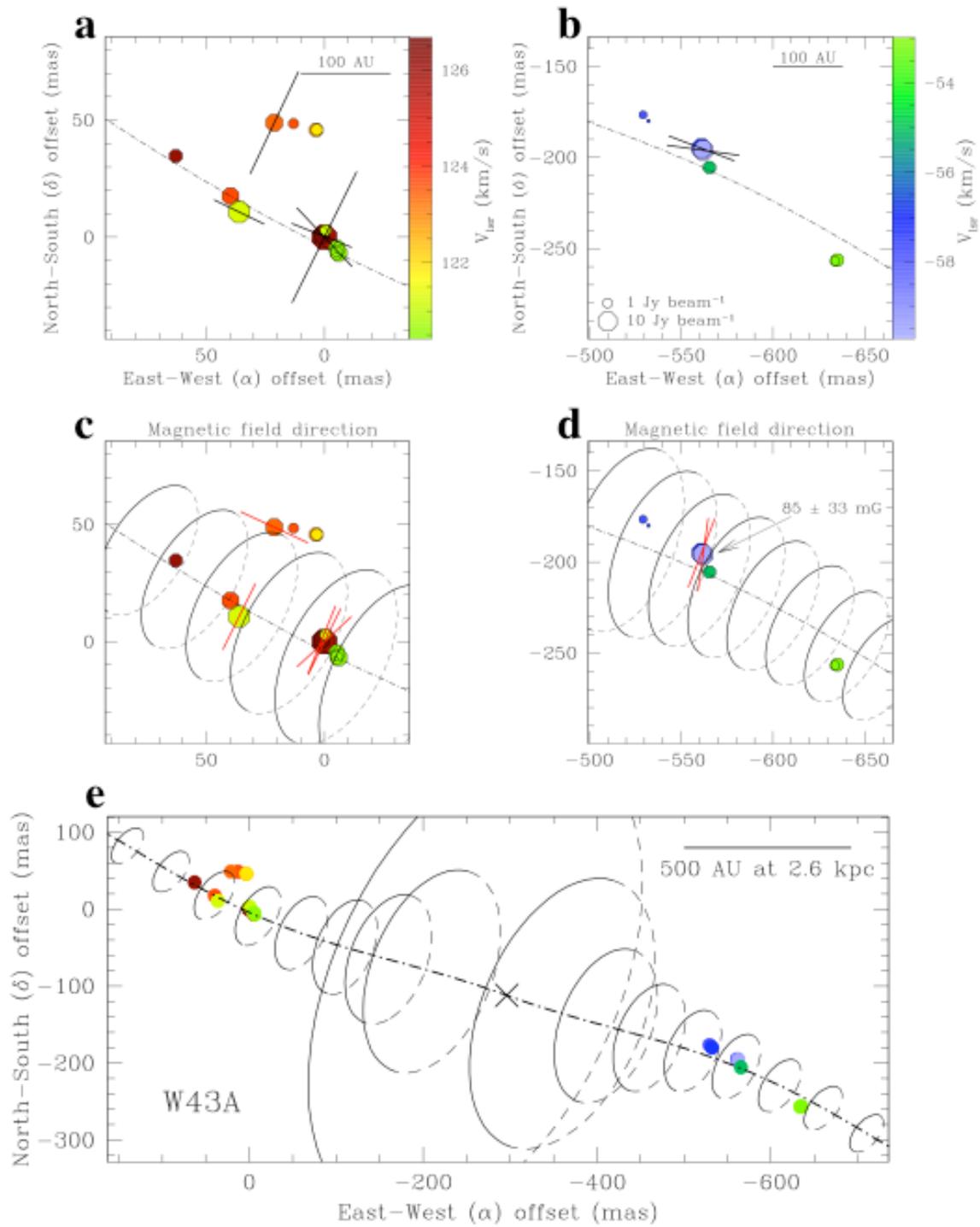

**Figure 1**: The spatial distribution, linear polarization vectors and inferred magnetic field direction of the H₂O masers in the jet of W43A. H₂O maser emission ($6_{16}$-$5_{23}$ line at 22.235080 GHz) was observed with the VLBA with an angular resolution of 0.5 milliarcsecond (mas) in right ascension ($\alpha$) and 1.0 mas in declination ($\delta$). In panel **a-d** the maser features are indicated by coloured

hexagons, scaled logarithmically according to their flux density in Jy beam$^{-1}$. The maser features in panel **e** are indicated by coloured dots. The colour denotes the radial velocity $V_{lsr}$ with respect to the local standard of rest (LSR). The dashed-dotted line indicates the model of the precessing jet of W43A ($v$ = 145 km s$^{-1}$, inclination 39°, position angle 65° east of north, 5° precession with a 55 year period). The cross in panel **e** denotes the location of W43A at approximately[27] $\alpha$(J2000)=18$^h$ 47$^m$ 41.166$^s$ and $\delta$(J2000)=-01°45'11.7". In panel **a** and **b** the logarithmically scaled linear polarization vectors are shown. In panel **c** and **d** we indicate the derived magnetic field direction with red vectors. The vectors reveal a toroidal magnetic field along the jet. The perpendicular magnetic field direction of the Northern-most maser feature indicates that these masers are likely located at the edge of the jet, where the magnetic field direction is along the projected jet direction due to shear between the compressed toroidal field lines inside and the uncompressed field lines outside the jet. This indicates a ~5° jet opening angle. The magnetic field strength derived from the circular polarization (Fig.2) is indicated in panel **d**. In panel **c-e** the ellipses show the toroidal magnetic field component along the jet. The sizes of the ellipses scale according to a $r^{-1}$ decrease of the magnetic field strength with increasing distance $r$ from the star.

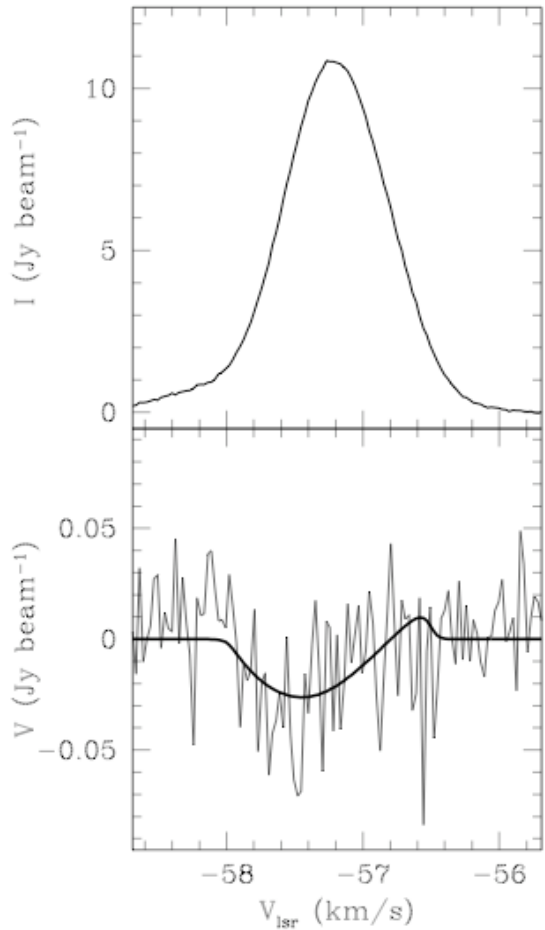

**Figure 2**: The total power (I) and circular polarization spectrum (V) of a 22 GHz $H_2O$ maser feature in the southern tip of the collimated jet of W43A. The spectra have a velocity resolution of 0.027 km s$^{-1}$ and a r.m.s. noise of 14 mJy beam$^{-1}$. The feature has a peak flux of 10.85 Jy beam$^{-1}$ and a radial velocity $V_{lsr}$=-57.19 km s$^{-1}$. The thick solid line in the lower panel is the best fit model to the circular polarization spectrum giving a circular polarization fraction $P_V$ = 0.33 ± 0.09 %. The s.d. error is determined from the r.m.s. noise on the total intensity and circular polarization spectra. The resulting magnetic field strength on the maser feature is $B_{||}$ = 85 ± 33 mG, where the increased s.d. error is the result of an added uncertainty due to the $H_2O$ maser radiative transfer model used in the fit.